\documentclass{iaus}
\usepackage{graphics}

  \checkfont{eurm10}
  \iffontfound
    \IfFileExists{upmath.sty}
      {\typeout{^^JFound AMS Euler Roman fonts on the system,
                   using the 'upmath' package.^^J}%
       \usepackage{upmath}}
      {\typeout{^^JFound AMS Euler Roman fonts on the system, but you
                   don't seem to have the}%
       \typeout{'upmath' package installed. iaus.cls can take advantage
                 of these fonts,^^Jif you use 'upmath' package.^^J}%
      }
  \else
  \fi

  \checkfont{msam10}
  \iffontfound
    \IfFileExists{amssymb.sty}
      {\typeout{^^JFound AMS Symbol fonts on the system, using the
                'amssymb' package.^^J}%
       \usepackage{amssymb}%

      }{}
  \fi

  \IfFileExists{amsbsy.sty}
    {\typeout{^^JFound the 'amsbsy' package on the system, using it.^^J}%
     \usepackage{amsbsy}}
    {}

\newsavebox{\astrutbox}
\sbox{\astrutbox}{\rule[-5pt]{0pt}{20pt}}

\newcommand\etal{\mbox{\textit{et al.}}}

\title[Outskirts of Galaxy Clusters: intense life in the suburbs]
      {ISOCAM Observations of Intermediate-redshift Galaxy Clusters at 7 and 15 $\mu$m}

\author[R. P\'erez-Mart\'inez   {\it et al.\/}]%
{R. P\'erez-Mart\'inez $^{1,2}$%
L. Metcalfe $^{1,2}$, D. Coia $^3$, A. Biviano $^4$,\break 
B. McBreen$^3$, B. Altieri $^2$, C. S\'anchez-Fern\'andez $^{1,2}$ 
}   

\affiliation{$^1$ISO Data Centre, European Space Agency, Villafranca del Castillo, P.O. Box 50727, 28080 Madrid, Spain.\\[\affilskip]
$^2$XMM-Newton Science Operations Centre, European Space Agency, Villafranca del Castillo, P.O. Box 50727, 28080 Madrid, Spain.\\ [\affilskip]
$^3$Physics Department, University College Dublin, Stillorgan Road, Dublin 4,
Ireland.\\ [\affilskip]
$^4$INAF/Osservatorio Astronomico di Trieste, via G.B. Tiepolo 11, 34131, Trieste,
Italy.\\ [\affilskip]
}

\pubyear{2004}
\volume{195}
\pagerange{1--8}
\date{?? and in revised form ??}
\jname{Outskirts of Galaxy Clusters: intense life in the suburbs}
\editors{A. Diaferio, ed.}
\begin{document}

\maketitle

\begin{abstract}
The gravitationally lensing clusters A370, A2218, A1689 and
CL0024+1654 were observed with the Infrared Space Observatory (ISO) using
ISOCAM at 6.7 $\mu$m and 14.3 $\mu$m (hereafter 7 $\mu$m and 15
$\mu$m respectively). 

A total of 178 sources were detected in the
whole set, 70 of them being cluster objects. The spectral energy
distribution of a subset of sources was calculated using GRASIL.
The results for the total infrared luminosity and the estimation of
the star formation rate are presented for the non stellar objects
for which the SED has been determined. The majority of the cluster
galaxies in A2218 are best fit by models of quiescent ellipticals. 
In Cl0024+1654, most of the galaxies lying on the Butcher-Oemler region of the
colour-magnitude diagram are best fit by disk galaxies, while those on the main 
sequence area have in general SEDs corresponding to post-starburst galaxies.

The population of each cluster is compared with the field population,
as well as with the population of other clusters. A significant number
of Luminous IR Galaxies (LIRGs) is detected in CL0024+1654, while
only one LIRG has been observed in total in A370, A1689, and
A2218. This result supports the link between LIRGs in clusters and
recent or ongoing cluster merger activity as well as the need for
extending the observations to the outer parts of clusters.
\end{abstract}

\firstsection 

\section{Introduction}

During the science operations of the Infrared Space Observatory
(ISO) (\cite[Kessler \etal\ \space 1996]{MFK96}), around 180 
kiloseconds were devoted by a number of observers to the
observation of several galaxy clusters exhibiting strong
gravitational lensing (\cite[Metcalfe \etal \space 
2003]{metcalfe2003}, Paper I hereafter, \cite[Altieri \etal \space
1999]{Altieri1999}, \cite[Barvainis \etal\space 1999]{barv99},
\cite[L\'emonon \etal\space 1998]{lem98}). 
These programmes mainly aimed to use lensing amplification to extend
the sensitivity of ISOCAM survey measurements. Paper I reported
field-source counts extending to the faintest levels achieved with
ISOCAM and complementing the field surveys (\cite[Elbaz \etal
\space 2002]{elb02}). 

The observations additionally provided extensive mid-infrared 
photometry of the clusters themselves. These results have been
combined with data at other wavelengths, available in the
literature. We have  focused on the two well known galaxy clusters:
A2218 and CL0024+1654.  The spectral energy distributions of the
cluster members are presented, as well as estimations of the
stellar formation rate for both clusters.

In Section 2 the data set is described. In Section 3 the main
results on A2218 are presented, while those of CL0024+1654 can be found
in Section 4. In Section 5 we compare the spectral energy
distributions and stellar formation rates found in the clusters with
those of the field galaxies.

We assume $H_0=70 \, km \, s^{-1} Mpc^{-1}$, $\Omega_0=0.3$ and
$\Omega_\Lambda=0.7$. With this cosmology the cluster luminosity
distances are 845 Mpc and 2140 Mpc, for A2218 and Cl0024+1654
respectively.  At the respective distances of A2218 and CL0024+1654, 1
arcsecond  corresponds to 2.97 kpc and 5.3 kpc and the age of the
Universe is 11.3 Gyr  and 9.3 Gyr.

\section{The Data}
A2218 and Cl0024+1654 were observed with ISOCAM on board ISO, using the
LW2 and LW3 filters (6.7$\mu$m and 14.3$\mu$m respectively,
hereafter 7$\mu$m and 15$\mu$m). In both cases the observations
were carried out in raster mode, to cover a large sky area and to 
reduce the effects of flat-fielding limitations. In the
case of A2218 the rasters were micro-scanned, i.e. the raster step
size included a fraction of a pixel distance to achieve better
spatial resolution.

The total observing time was 6.2 hours per filter for both
clusters, except for the 7$\mu$m measurement on Cl0024+1654,
which  used only a little over 3000 seconds.  The total area covered 
in A2218 was 20.5 arcmin$^2$ while
in Cl0024+1654 areas of 28.6 and 37.8 arcmin$^2$ were observed at 
7$\mu$m and 15$\mu$m respectively.

For further explanation of the observations and the data reduction
refer to Paper I, \cite[Biviano \etal \space 2004]{biviano04} (hereafter 
Paper II), and \cite[Coia \etal \space  2004]{Coia04}
(hereafter Paper III).   

\section{Mid-infrared cluster sources in A2218}

A2218 is a massive galaxy cluster at $z=0.175$. Its $R_{vir}$ is
1.63 $h^{-1}Mpc$ and its $M_{vir}$ is 18.27 $h^{-1}10^{14} M_\odot$.
In total 76 sources were detected in the A2218 field
(see Paper I and II).

The velocities and spatial distribution within the cluster
of these MIR selected objects do not differ from the rest of the
optically selected sources in A2218, so they do not represent a
special population from a dynamical point of view.

Photometric data were collected from the literature to calculate
spectral energy distributions of the sample. Ten different model
SEDs were computed with GRASIL (\cite[Silva \etal \space ,
1998]{silva98}) and used to find a best fit model for the observed
cluster members. This code calculates the spectral evolution of
galaxies by taking into account the effect of dust distributed in
the environment as AGB envelopes, diffuse interstellar medium and
molecular clouds.

The model SEDs used were distributed as follows: \textbf{E:} Three
models of early type galaxies with initial bursts of star formation
lasting 0.5, 1.0 and 2.0 Gyr.  These models were used by
\cite[Granato \etal \space  (2001)]{granato01} and reproduce ellipticals and S0s in the nearby
Universe. \textbf{S:} Three models of disk galaxies characterized
by different values of the gas infall time-scales and the
efficiency term in the Schmidt-type law. \textbf{SB:} Two
starburst models adjusted to fit the moderate starburst galaxy M82
and the strong starburst galaxy Arp220. The starburst is
characterized by an e-folding time of 0.05 Gyr and involves
$\approx 0.01$ and $\approx 0.1$ of the total mass of the galaxy
respectively. \textbf{PSB:} Two post-starburst models corresponding
to the previous ones, observed 1 Gyr after the event. All these 10
models were computed at three different ages, forming a final
set of 30 possibilities.

The observed and model SEDs were compared using a standard $\chi^2$
procedure, having as free parameters the model SED and the flux
normalization. Most of the objects presented SEDs typical of
early-type galaxies with none or very small star formation
activity. Only four of them presented SFRs above 1 $M_\odot
yr^{-1}$ (see Paper II). The rest were best fit by SEDs
of passively evolving galaxies. Considering the V-I vs I-K diagram
used in \cite[Smail \etal \space  2001]{smail01}, the more luminous
(and redder) of the cluster members are 5 to 10 Gyr old.

The median infrared luminosity $L_{IR}$ of the cluster sources is
$6\times10^8 L_\odot$, and none of the objects are significantly
blue. This indicates extremely mild star
formation activity if any. None of the MIR selected galaxies lie in
the Butcher-Oemler region of the V-I vs I diagram. 

Several independent analysis based on optical and X-ray
observations suggest than A2218 is not a dynamically relaxed
cluster. In any case the cluster dynamical status does not appear to
affect the MIR properties of its members in the region surveyed. 

For further explanation of the data analysis and discussion
refer to Paper II.

\section{Mid-infrared cluster sources in Cl0024+1654}

Cl0024+1654 is a rich cluster of galaxies at $z = 0.395$ with a
spectacular system of gravitationally lensed arcs though, unlike
other clusters with these arc-like features, it harbours no central
cD galaxy. Its $R_{vir}$ is 0.94 $h^{-1}Mpc$ and its $M_{vir}$ is
6.42 $h^{-1}10^{14} M_\odot$. It is one of the two clusters
studied by Butcher and Oemler in 1978 and has a significant
fraction of blue galaxies ($f_B=0.16$). As reported by \cite[Treu
\etal \space  2003]{treu03}, its core is composed of 73\% 
early-type galaxies, with this proportion decreasing to 50\% at 1 Mpc
and to 43\% at its periphery at about 5 Mpc.

Thirteen confirmed cluster sources were detected on the ISOCAM
$15\mu m$ map of this field, above the $4\sigma$ significance level
(Paper III). They all show spatial distributions and velocity
characteristics in agreement with the optically selected cluster
members. They have special characteristics from the dynamical point
of view. 

The spectral energy distributions of these sources were obtained
following the same strategy as in the A2218 case, though only two
different ages of each model were considered. Good fits were
obtained for the 12 objects with available optical data. Their
total infrared luminosities were calculated based on these SEDs,
yielding values ranging from $4.7 \times 10^{10}L_\odot$ to $4.5
\times 10^{11}L_\odot$. Six of these twelve sources have total
infrared luminosities above $10^{11} L_\odot$, classifying them as
Luminous Infrared Galaxies (LIRGs), and 4 are well within $1\sigma$ 
of this luminosity threshold. 54\% are in the Butcher-Oemler region 
in a colour-magnitude diagram, and these are best fit by spiral
galaxy models. On the other hand, the main sequence objects are best
fit by post-starburst galaxies (see Paper III).

The results show that all thirteen cluster objects detected at
15$\mu m$ present traces of significant star formation activity.
The star formation rate can be derived from the luminosity of
the [OII] line, according to the relation SFR[OII] $\approx 1.4
\times 10^{-41}$ L[OII] $M_{\odot} yr^{-1}$ (\cite[Kennicut
1998]{kennicut98}), though this method heavily depends on dust
extinction, and on the metallicity and ionization of the medium. The SFR
can also be obtained from the global infrared luminosity according to the
relation $SFR[IR] \approx 1.71 \times 10^{-10}L_{\odot}$
(\cite[Kennicut, 1998]{kennicut98}). The mean value of the SFR
obtained for the cluster members is $30 M_\odot yr^{-1}$.  For three
of the sources the SFR has been comparatively calculated through 
the [OII] flux and through the $L_{IR}$ value. The mean value obtained 
through the [OII] flux is 10 times lower than that obtained with 
the $L_{IR}$ value (Paper III). This indicates that the stellar formation
activity may be heavily obscured by dust in some or all the
observed objects, as also found in A1689 (see \cite[Duc \etal
\space  2002]{duc02}).

Comparing these results with what has been obtained for other
clusters requires a scaling of the LIRG count to allow for 
different cluster virial radii, masses, distances, and the sky 
area scanned in each case. For details on this scaling refer 
to Paper III. In A370 (a cluster similar to Cl0024+1654 in 
distance, mass and size) only one cluster object is detected 
at 15 $\mu m$. This sole source is a LIRG, while something like
10 would be expected to have been observed by analogy with the
Cl0024+1654 case. The difference may be due to the dynamical 
characteristics of Cl0024+1654. According to
\cite[Czoske \etal \space  2002]{czoske02}, the cluster is
undergoing a major merging event, which may be responsible for the
high SFR and $L_{IR}$ observed in some of its galaxies. Yet when 
comparing with A1689, a cluster similar to Cl0024+1654 in mass 
and size, but at redshift 0.18, and also having an un-relaxed 
dynamical status, it is found that the instances of very high
SFR in Cl0024+1654 are much more than found in
A1689, which nonetheless, as mentioned above, shows elevated SFR 
(\cite[Duc \etal \space  2002]{duc02}). In this case, the 
portion of the cluster surveyed plays a
major role, since in the case of A1689 the scanned area belongs to
the inner part of the cluster (0.5 Mpc from the centre), while in
Cl0024+1654 the area observed has a cluster-centric radius of 
2 Mpc, extending to the outer region where the BO galaxies lie.

\section{Field Galaxies}

We have also obtained SEDs of a sample of 14 of the field galaxies
detected by ISOCAM in these areas. In general, the field galaxies
show stellar formation activity partially or mostly obscured by
dust, although this activity is not extremely high in none of them.
They are well fitted by models with no significant AGN
contribution, being mainly normal spirals and post-starburst
galaxies with no major activity events in the last 1 Gyr. 60\% of
these objects classify as LIRGs, with a median star formation
rate calculated from the $L_{IR}$ of 22 $M_\odot yr^{-1}$ (Paper
II). These values are significantly higher than the ones obtained
in A2218, though are comparable to the ones in Cl0024+1654. The
redshift distribution of the field galaxies ranges from 0.1 to 1.1,
with a median of $z = 0.6$.

\section{Summary and Conclusions}

We discussed the ISOCAM observations of a number of galaxy clusters,
concentrating mainly on the results for A2218 and Cl0024+1654. In the 
first of these it is found that most of the cluster members detected 
at 7$\mu$m and 15$\mu$m have SEDs resembling early-type galaxies with 
little or no star formation activity. None of them qualify as Butcher-Oemler
galaxies and they have a median $L_{IR}$ of $6 \times 10^8 L_\odot$,
far below the LIRG luminosity threshold. If the cluster has a 
non-relaxed dynamical status, this does not seem to affect the MIR properties
of the cluster galaxies within the area studied. 

The cluster Cl0024+1654 harbours 13 sources detected at 
15$\mu$m, 11 of them LIRGs having MIR luminosities within 10\% of
the LIRG threshold of $1 \times 10^{11} L_\odot$, and six falling above that
threshold. Six out of the eleven for which I and R band data were available
lie in the BO region on a I-R vs I diagram. The SFR calculated through the
$L_{IR}$ values are consistent with disk-like and post-starburst
galaxies, and are an order of magnitude higher than the values obtained
through the equivalent width of the [OII] spectral line. This may
be due to heavy absorption of the UV and visible photons by dust,
and their subsequent re-emission in the MIR spectral range.

Comparison with other clusters and with field galaxies suggests a
link between the dynamical history of a cluster and star formation in 
its constituent galaxies.  But this link may or may not be observed, depending
on which part of the cluster is surveyed.  The highest number of LIRGs and 
the highest SFR are found in the
cluster which is undergoing a merging event and the outskirts of which
have been included in the observations. It is necessary to extend
these studies in number of targets studied and in area covered in order to
reach firmer conclusions. In this sense, SPITZER can provide excellent 
data for this purpose.


\end{document}